\documentclass[a4paper,12pt]{article}
\usepackage{amsmath}
\usepackage{pstricks}
\usepackage{pst-node}
\usepackage[ansinew]{inputenc}
\usepackage{amssymb,amsmath}
\usepackage{amsfonts}
\usepackage{epsfig}

\def\a{\alpha}

\def\g{\gamma}

\def\o{\omega}

\font\Sets=msbm10

\def\Integer {\hbox{\Sets Z}}    

   \def\Natural {\hbox{\Sets N}}

\def\be{\begin{equation}}       \def\ba{\begin{array}}

\def\ee{\end{equation}}         \def\ea{\end{array}}

\def\bea {\begin{eqnarray}}      \def\eea {\end{eqnarray}}

\def\bean{\begin{eqnarray*}}    \def\eean{\end{eqnarray*}}

\def\RA {\ \Rightarrow\ }

\def\<{\langle} \def\({\left(}  \def\>{\rangle} \def\){\right)}

\newtheorem{exi}{Example}

\author{Elena Kartashova\\ RISC, J. Kepler University\\
 Linz, Austria\\  E-mail: lena@risc.uni-linz.ac.at}
\title{Fast Computation Algorithm for Discrete Resonances among Gravity Waves\footnote{
Author acknowledges support of the Austrian Science Foundation (FWF)
under projects SFB F013/F1304.}}

\date{}
\begin{document}
%generates titel
\maketitle
%\tableofcontents {
\section{Introduction}

Classical theory of wave turbulence was  developed  for description
of statistic properties of weakly nonlinear waves in infinite
domains. Importance of resonant interactions among these waves
 was first pointed out in 1961 by O.Phillips \cite{phil61} and this led to construction of the first
kinetic equation \cite{has}. Later on it was established by many
researchers that effects of finite length are not described by the
classical theory and kinetic equations, and 20 years later, in 1981,
O.Phillips wrote that "new physics, new mathematics and new
intuition is required" \cite{phil85} in order to understand
energetic behavior of these systems. Discrete effects appear when
waves "notice" boundary conditions, i.e. wave lengths are comparable
to the sizes of interaction domain, and therefore wave numbers are
integers. First theoretical result on the behavior of discrete waves
systems was published in 1990 in \cite{PHD1} where main distinction
between continuous and discrete wave systems was established:
stochastic interactions between all waves (infinite domain) {\it
versus} interactions within small independent groups of waves and
existence of non-interacting waves (finite domain). Theory of
discrete wave systems was then presented in \cite{PRL}, \cite{AMS},
\cite{Clip} thus giving a rise to the following qualitative picture
of wave turbulence: short waves are described by power energy
spectra and kinetic equations, long waves are described by Clipping
method and dynamic equations. Numerous papers of last few years
showed that this qualitative picture has to be modified because some
discrete effects (frozen turbulence, mesoscopic turbulence, etc.)
are also observable in the short-waves' part of the wave spectra
(\cite{zak3}, \cite{zak4}, \cite{naz} and others). Discovery of
intermittent patterns in that part of wave spectra which is supposed
to be described by a kinetic equation only, gave us the reason to
revisit the very foundations of wave turbulence theory. It turned
out that its basic theoretical ground - KAM theorem - is proven {\bf
not for all} short waves, a countable number of short waves with
rationally connected frequencies is excluded from consideration
leaving some "gaps" in the wave spectrum. These gaps correspond to
the waves whose ratio of frequencies is algebraic number of degree
$\le 2$, for instance, a rational number which is usually understood
as necessity for the corresponding dispersion function to be a
rational function. A model of laminated turbulence was recently
presented \cite{lam} which consists of two layers, discrete one (in
the whole range of wave numbers) and continuous one (in the
short-waves' part of the wave spectrum). The main fact allowing to
construct this model is following: frequencies of discrete
resonantly interacting waves are rationally connected though the
{\bf dispersion function itself can be highly irrational}. Thus, the
model of laminated wave turbulence "fills" the gaps left by KAM
theory and explains in particular coexistence of power energy
spectra and coherent structures in the
short-waves' part of wave spectrum.\\

Model of laminated turbulence brings our attention to the new
question: how to compute discrete resonances in large computational
domain? Indeed, say for 4-waves interactions of 2-dimensional
gravity waves, the resonance conditions can be regarded in the form
$$ \sqrt{k_1}+\sqrt{k_2}=\sqrt{k_3}+\sqrt{k_4},\quad
\vec{k}_1+\vec{k}_2=\vec{k}_3+\vec{k}_4 $$ where
$\vec{k}_i=(m_i,n_i), \ \forall i=1,2,3,4$ and
$k_i=|\vec{k}_i|=\sqrt{m_i^2+n_i^2}$. It means that in a finite but
big enough domain of wave numbers, say $|m|,|n| \le 1000$, direct
approach leads to necessity to perform computations with integers of
order $10^6$. These computations in a substantially smaller domain
$|m|, |n|\le 128$ took 3 days with Pentium-4 (\cite{naz1}).\\

The main goal of this paper is to show that constructions of
independent classes of resonantly interacting waves \cite{AMS} can
be re-formulated as a pure mathematical procedure which allows to
reduce drastically computation time for this sort of equations.
Gravity waves are taken as our main example. Some necessary
 results on discrete resonances of gravity waves are given in \cite{AMS} but
formulations there are too formal and partly inaccurate.
 We re-formulate all necessary
results on construction of classes for these waves in Sec.2 and give
some illustrative examples. In Sec.3 we describe the computation
algorithm and some preliminary results of numerical simulation.
Brief discussion is given in Sec.4.

\section{Construction of classes}
In this section we regard 4-waves resonances of 2-dimensional
gravity waves in the form \be \label{grav_res}
\sqrt{k_1}+\sqrt{k_2}=\sqrt{k_3}+\sqrt{k_4}\ee \be\label{grav_lin}
\vec{k}_1+\vec{k}_2=\vec{k}_3+\vec{k}_4. \ee  To construct
independent classes of resonantly interacting we need following
definition.

\paragraph {Definition} Let $\vec{k}=(m,n)$ be a vector with integer
coordinates, $m,n \in \Integer$. Represent the square root of the
norm $|\vec{k}|$ as
$$
(m^2+n^2)^{1/4}=\g q^{1/4}, \ \ \g,q \in \Natural
$$
and $q$ does not contain fourth degrees:
$$
q=p_1^{\a_1} \cdot \cdot \cdot p_s^{\a_s}, \ \ \a_i \in \{1,2,3\},
$$
with all different prime $p_i$. Number $q$ is called  index of a
vector $\vec{k}=(m,n)$ and number $\g$ is called its weight.
Class $Cl_q$ of index $q$ is set of all vectors with this index.\\

For instance, vectors $(1,3)$ and $(12,4)$ belong to the same class
$Cl_{10}$ of index 10, with weights $\g=1$ and $\g=2$
correspondingly, while vector $(2,1)$ lies in $Cl_5$ of index 5,
with weight $\g=1$.

\paragraph {Lemma} Let vectors $\vec{k}_1, \vec{k}_2, \vec{k}_3,
\vec{k}_4$ construct a solution of (\ref{grav_res}), then only two
cases are possible:

Case 1: all  vectors belong to the same class,

Case 2: all vectors belong to two different classes $Cl_{q_1},
Cl_{q_2}$ in such a way that there exist $q_1,q_2:$
$$
\vec{k}_1, \vec{k}_3 \in Cl_{q_1} \quad \mbox{and} \quad \vec{k}_2,
\vec{k}_4 \in Cl_{q_2}
$$
or
$$
\vec{k}_1, \vec{k}_4 \in Cl_{q_1} \quad \mbox{and} \quad \vec{k}_2,
\vec{k}_3 \in Cl_{q_2}.
$$
In Case 2 all solutions are symmetric, i.e. vectors belonging to
each class must have the same norm. All asymmetric solutions, if
any, are described by Case 1.\\

The statement of Lemma follows immediately from elementary
properties of integers and we are not going to the detailed proof
here. The general idea of the proof is very simple indeed: two
different irrational numbers can not satisfy any equation with
rational coefficients. For instance, equation $
a\sqrt{3}+b\sqrt{5}=c $ has no solutions for arbitrary  rational
$a,b,c$. It means that irrationalities corresponding to classes
indices in (\ref{grav_res}) have to be rid off in order to construct
its integer solutions. In general,  one to four different
irrationalities can appear in (\ref{grav_res}), simple consideration
show that all cases but two described below
 give no integer solutions.\\

In Case 1, all four irrationalities are equal and can be canceled,
so that (\ref{grav_res}) is reduced to
$$
\g_1q^{1/4}+\g_2q^{1/4}=\g_3q^{1/4}+\g_4q^{1/4} \RA
\g_1+\g_2=\g_3+\g_4.
$$
  Let us regard as example a couple of
asymmetric solutions
 given in \cite{naz}: $$ (-4,0), (49,0), (9,0), (36,0) \quad
\mbox{and} \quad (-20,15),(-20,-15),(-49,0),(9,0).$$ An easy check
shows that all 8 vectors belong to the same class $Cl_1$ of index 1,
with weights $\ \g_1=2, \g_2=7, \g_3=3, \g_4=6 \ $ and $ \ \g_1=5,
\g_2=5, \g_3=7, \g_4=3 \ $ correspondingly, so that condition $ \
\g_1+\g_2=\g_3+\g_4 \ $ is full-filled in both cases. It is
important to understand that  4 vectors constructing some asymmetric
 solution must belong to the same class but not necessarily
 to $Cl_1$ as in the solutions above.
 In the next section
 some asymmetric solutions are presented which lie in other classes. \\

In Case 2, the irrationalities with corresponding coefficients in
front of them must be pair-wise equal to be canceled, for instance
$$
\g_1q_1^{1/4}+\g_2q_2^{1/4}=\g_3q_1^{1/4}+\g_4q_2^{1/4} \RA
\g_1=\g_3 \quad \mbox{and} \quad \g_2=\g_4.
$$

Notice that Lemma gives only {\bf necessary} condition of the
existence of a solution. This means that to find a solution we have
first to construct classes, then find solutions within the
corresponding classes and check linear conditions afterwards.
Obviously,  (\ref{grav_res}),(\ref{grav_lin}) have infinitely many
solutions - for instance, quartets  \be \label{trivial}
\vec{k}_1=(a,b),\ \ \vec{k}_2= (c,d),\ \ \vec{k}_3= (a,b), \ \
\vec{k}_4=(c,d) \ee and \be \label{sym}\vec{k}_1=(a,b),\ \
\vec{k}_2= (c,a-b+c),\ \ \vec{k}_3= (b,a), \ \ \vec{k}_4=(a-b+c,c)
\ee with arbitrary integers $a,b,c,d$ give its solution, as well as
any  proportional quartets corresponding to multiplication of all
wave-numbers on the same integer.  Less trivial example of
"tridents" \be \label{trident} \vec{k}_1=(a,0),\ \
\vec{k}_2=(-b,0),\ \ \vec{k}_3=(c,d),\ \ \vec{k}_4=(c,-d) \ee
possess two-parametric series of solutions \cite{naz}: \be
\label{ser}a=(s^2+t^2+st)^2, \ b=(s^2+t^2-st)^2, \ c=2st(s^2+t^2), \
d=s^4-t^4 \ee with arbitrary integer $s,t$. This series gives
solutions of both (\ref{grav_res}),(\ref{grav_lin}) though perhaps
not all of them. Parametrization (\ref{ser}) is constructed in such
a way that norms of all four vectors are full squares, i.e. again
vectors $ \ \vec{k}_1, \ \vec{k}_2, \ \vec{k}_3, \ \vec{k}_4 \ $
belong to the same class $Cl_1$ of index 1, with weights
$$
\g_1=s^2+t^2+st,\ \  \g_2=s^2+t^2-st,\ \  \g_3=\g_4=s^2+t^2,
$$
providing $\g_1+\g_2=\g_3+\g_4$ for any parameters $s,t$.\\

 There is no known way to construct general solution of
(\ref{grav_res}),(\ref{grav_lin}) and even a construction of some
particular solutions' series  demands a lot of skillful work and a
bit of luck. On the other hand, this hard work is mostly not
well-paid because coming back to physical problem setting, we
usually need {\bf all solutions} in some spectral domain, not only
those which could be nicely parameterized. In the next section we
use Lemma to construct fast computer algorithm for finding solutions
of (\ref{grav_res}),(\ref{grav_lin}) which is  a challenging
numerical problem.

\section{Scheme of numerical algorithm}
 As
it was shown in previous section, equations
(\ref{grav_res}),(\ref{grav_lin}) have infinitely many solutions.
But not all of them are interesting from physical point of view. We
follow \cite{pok} in classification (written out up to the cyclic
change of indices) of all possible solutions according to their role
in energy transfer : (I) trivial resonances $\vec{k}_1=\vec{k}_3, \
\vec{k}_2=\vec{k}_4$ which do not redistribute the energy, an
example is given by (\ref{trivial}); (II) symmetrical resonances
$|\vec{k}_1|=|\vec{k}_3|, \ |\vec{k}_2|=|\vec{k}_4|$ which do not
generate new wavelengths and therefore  do not transfer the energy
through the scales, an example is given by (\ref{sym}); (III)
asymmetrical resonances playing major role in energy transfer, an
example is given by (\ref{trident}). In this paper we are primary
interested in asymmetric solutions which means in terms of Lemma
 that all vectors belong to the same class
(Case 1). Not going into all programming details we present here
just underlying
main ideas of our algorithm and first results of computer simulations. \\

{\it Step 1.}  Create first auxiliary array of primes $A_p$ in the
given spectral domain  $p \le D$  using   Eratosthenes' Sieve
procedure which eliminates composite numbers from the list of
natural numbers and can be briefly described as follows:
\begin{itemize}
\item{} Create an array $A_d$ containing all integers $1,...,d$,
write number "1" into array $A_p$, mark "1" in the array $A_d$ as
used element;
\item{} Introduce cycle variable $i=1,..., \sqrt{d}$;
\item{} Find first number in $A_d$ which is greater than 1 and is
not marked yet as a composite, denote it as $r$ and mark all numbers
$2r, \ 3r, \ 4r, \ ...$ as composite (at the first step $r=2$, at
the second $r=3$ and so on);
\item{} write number $r$ into array of primes $A_p$;
\item{} put $i=r$ and repeat the procedure.
\end{itemize}

{\it Step 2.} Create second auxiliary array $A_q$ of  possible
indexes constructed of the primes obtained at the Step 1. Indexes
are computed directly by formula
$$
q=p_1^{\a_1} \cdot \cdot \cdot p_s^{\a_s}, \ \ \a_i \in \{1,2,3\},
$$
in the domain $q \le 2^{1/4}\sqrt{d} < 1.19 \sqrt{d}.$ Say, for
$d=1000,$ we have $q < 37.63$, i.e.
it is enough to compute for $q \le 37.$\\

{\it Step 3.} Find all integer solutions of the linear equation
$$\g_1+\g_2=\g_3+\g_4$$
in the domain  $\g_i \le 2^{1/4}\sqrt{d}, \quad \forall i=1,..4.$\\

{\bf Remark.} Further it will be necessary to check when expressions
under the radicals $ \g^4 q $ can be represented as a sum of two
squares. Here one has to make use of the well-known Euler theorem:
representation of a positive number as a sum of two squares which is
possible {\bf iff} each of its prime factors  of the form $4t+3$
occurs as an even power. For instance,  pair $\g=3, \ q=1$ has 3 in
even degree,  expression under the integral is $3^4 \cdot 1 =81$,
representation as a sum of two squares exists and number
$81=9^2+0^2$ should be written into the array $A_{\g, q}$. On the
other hand, for pair $\g=2, \ q=3,$ the expression under the radical
has factor 3 in the odd power, i.e. number 48 can not be represented
as a sum of two squares  and it is not an element of the array
$A_{\g, q}$. Notice that all prime factors of $\g$ have even powers
which means that only prime factors of $q$ has to be investigated.
Consequently, an arbitrary presentation of $q$ as a sum of two
squares, multiplied
by $\g^4,$  gives corresponding presentation of the expression under the radical.\\

{\it Step 5.} Check Euler theorem for all elements of  $A_q$ and
construct array $\tilde{A}_q$ with "allowed" elements only; find all
presentations as a sum of two squares for each "allowed" element. Obviously, one number
can be decomposed into sum of two squares in more then one way, for instance
$1105=4^2+33^2=9^2+32^2=12^2+31^2=23^2+24^2.$\\

{\bf Remark.} Due to  Lemma, the search of 2-square-representations
for the numbers  $\le 2d^2$ is reduced  to the numbers $< 1.19
\sqrt{d}$. In particular, for $d=1000$ we have to compute these
presentations for the numbers in the domain $\le 37$ which can be
done by direct enumeration. The number $S$ of different
2-square-presentations of an integer is proportional to   the
difference between its prime factors of the form $4t+1$ and $4t+3$
(Lagrange
theorem).\\

  {\it Step 6.} For each $q \in \tilde{A}_q$ and each its possible
  2-square-presentation, $q=\tilde{m}_j^2+\tilde{n}_j^2, j=1,2,...,S$, compute
  all under-integral-expressions $\g_i^4(\tilde{m}_j^2+\tilde{n}_j^2)$
and check linear conditions on $m_j,n_j$:
$$
m_1+m_2=m_3+m_4,
$$
$$
n_1+n_2=n_3+n_4,
$$
with $m_1=\g_1^2 \tilde{m}_j, \  n_1=\g_1^2 \tilde{n}_j, \ $
$m_2=\g_2^2 \tilde{m}_j, \  n_2=\g_2^2 \tilde{n}_j, \ $
$m_3=\g_3^2
\tilde{m}_j, \  n_3=\g_3^2 \tilde{n}_j, \ $
$m_4=\g_4^2 \tilde{m}_j, \  n_4=\g_4^2 \tilde{n}_j. \ $ \\

Summarizing, Lemma allows to reduce domain of integers under
consideration from $2d^2$ to $2^{1/4}\sqrt{d}$ while classical
procedure (Eratosthenes' Sieve) and some known number-theoretical
results (Euler and Lagrange theorems) allow to further reduce the
computation time which is about 4 min. at Pentium-4 for spectral
domain $m_i, n_i \le 1000.$ In this domain we have found five
 asymmetric solutions which are not tridents:
$$\vec{k}_1=(495,90), \ \vec{k}_2=(64, 128), \ \vec{k}_3=(359, 118), \ \vec{k}_4=(200, 100),$$
$$\vec{k}_1=(675,225), \ \vec{k}_2=(64, 192), \ \vec{k}_3=(479, 237), \ \vec{k}_4=(260, 180),$$
$$\vec{k}_1=(810,45), \ \vec{k}_2=(128, 192), \ \vec{k}_3=(598, 117), \ \vec{k}_4=(340, 120),$$
$$\vec{k}_1=(855,360), \ \vec{k}_2=(64, 256), \ \vec{k}_3=(599, 356), \ \vec{k}_4=(320, 260),$$
$$\vec{k}_1=(990,180), \ \vec{k}_2=(128, 256), \ \vec{k}_3=(718, 236), \ \vec{k}_4=(400, 200).$$
These solutions belong to classes $\ Cl_5, \ Cl_{10}, \ Cl_{13}, \
Cl_{17}, \ Cl_{20} \ $ correspondingly, with weights $ \ \g_1= 15, \
\g_2=8, \ \g_3=13, \ \g_4= 10 \ $ for all solutions.\\

\section{Brief discussion}
We described here first version of our algorithm in order to show
how to use Lemma for reducing drastically computation time. At
present, this algorithm is written only for positive integer wave
numbers $m_i,n_i \in \Natural $. Using some simple symmetrical
considerations, one can adapt it for the case of arbitrary $m_i,n_i
\in \Integer. $ Another simple modification can be made in order to
find solutions for the case
$$
\sqrt{k_1}=\sqrt{k_2}+\sqrt{k_3}+\sqrt{k_4}, \quad
\vec{k}_1=\vec{k}_2+\vec{k}_3+\vec{k}_4$$ or for the case of 5-wave
interactions and more (of course, Lemma should be re-formulated).\\

Moreover, basing on this algorithm, one can develop  {\it generic
algorithm} for fast computation of discrete resonances among
different types of waves with dispersion function $\o$ being {\it an
arbitrary} polynomial function of the norm of wave vector,
$\o=\o(k).$ Different dispersion functions in case of two
dimensional waves will have different formula for index (Step 2 of
present algorithm); in case of three dimensional waves, i.e.
$k=|\vec{k}|=\sqrt{m^2+n^2+l^2},$ Lagrange and Euler theorems should
be replaced by some known number-theoretical results on the
decomposition of an integer into the sum of three squares. All other
algorithmic steps will be the same as above.  Development of such a
generic algorithm for a big class of
dispersion functions is our current object of interest. \\

Author is very grateful to Sergey Nazarenko for valuable discussions
and remarks.

\end{document}